\documentclass[lettersize,journal]{IEEEtran}

\usepackage[utf8]{inputenc}
\usepackage[T1]{fontenc}
\usepackage{amssymb,amsmath,amsfonts,amsthm,dsfont}
\usepackage{units}
\usepackage{graphicx}
\usepackage{float}
\usepackage{caption}
\usepackage{subcaption}
\usepackage{multirow}
\usepackage{physics}
\usepackage{hyperref}
\usepackage{enumitem}
\usepackage{orcidlink}

\ifCLASSOPTIONcompsoc
  \usepackage[nocompress]{cite}
\else
  \usepackage{cite}
\fi

\newcommand{\isep}{\mathrel{{.}\,{.}}\nobreak}

\newtheorem{theorem}{Theorem}
\newtheorem{proposition}[theorem]{Proposition}

\newtheorem{lemma}{Lemma}

\newtheorem{definition}{Definition}

\begin{document}

\title{Low-rank quantum state preparation\thanks{© 2023 IEEE. Personal use of this material is permitted. Permission from IEEE must be obtained for all other uses, in any current or future media, including reprinting/republishing this material for advertising or promotional purposes, creating new collective works, for resale or redistribution to servers or lists, or reuse of any copyrighted component of this work in other works.
}}

\author{
        Israel F. Araujo\orcidlink{0000-0002-0308-8701},
        Carsten Blank\orcidlink{0000-0003-3450-0823},
        Ismael C. S. Araújo \orcidlink{0000-0001-6240-0235},
        Adenilton J. da Silva\orcidlink{0000-0003-0019-7694}%
\IEEEcompsocitemizethanks{

\IEEEcompsocthanksitem I.F. Araujo is with Department of Statistics and Data Science, Yonsei University, Seoul, Republic of Korea.
\IEEEcompsocthanksitem C. Blank is with Data Cybernetics, 86899, Landsberg am Lech, Germany.
\IEEEcompsocthanksitem I.C.S. Araújo and A.J. da Silva are with Centro de Informática, Universidade Federal de Pernambuco, Recife, Pernambuco, Brazil.
}
}

\maketitle

\begin{abstract}
Ubiquitous in quantum computing is the step to encode data into a quantum state. This process is called quantum state preparation, and its complexity for non-structured data is exponential on the number of qubits. Several works address this problem, for instance, by using variational methods that train a fixed depth circuit with manageable complexity. These methods have their limitations, as the lack of a back-propagation technique and barren plateaus. This work proposes an algorithm to reduce state preparation circuit depth by offloading computational complexity to a classical computer. The initialized quantum state can be exact or an approximation, and we show that the approximation is better on today's quantum processors than the initialization of the original state. Experimental evaluation demonstrates that the proposed method enables more efficient initialization of probability distributions in a quantum state.
\end{abstract}


\section{Introduction}

\begin{figure}[ht]
    \centering
    \includegraphics[width=1.0\columnwidth]{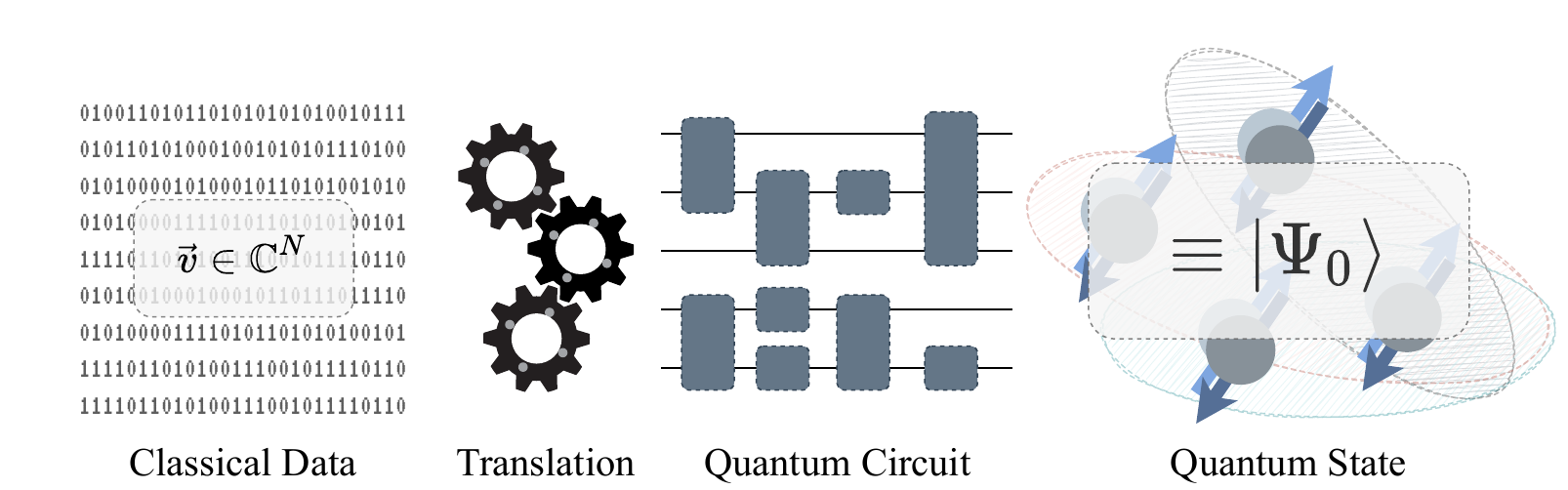}
    \caption{The process of state preparation. Classical data is converted into vector form and a translation process encodes the information in operations performed on the quantum computer, commonly encoded by a circuit. Executing this circuit initializes the state $\ket{\Psi_0}$.}
    \label{fig:state_preparation}
\end{figure}

\IEEEPARstart{Q}{uantum} devices can execute information processing tasks that classical computers cannot perform efficiently~\cite{arute2019quantum}. In some instances, this leads to exponential advantages in solving systems of linear equations~\cite{harrow2009quantum} and principal component analysis~\cite{lloyd2014quantum}. Additionally, there are known advantages in Monte-Carlo sampling~\cite{10.1098/rspa.2015.0301,10.1103/physreva.98.022321} in which a squared increase of convergence can be attained. Furthermore, quantum machine learning applications~\cite{biamonte2017quantum, schuld2018supervised, 10.1038/s41534-020-0272-6} may exhibit heuristic advantages. For all these applications, the initialization of a $n$-qubit quantum state, commonly called quantum state preparation (Fig.\ref{fig:state_preparation}), is an important step in quantum information processing. Encoding a $N$-dimensional (complex) vector requires $n$-qubits ($N=2^n$) and quantum circuits with $O(2^n)$ controlled-NOT (CNOT) gates~\cite{Aaronson2015,Leymann}.
Therefore, several works focus on the development of algorithms that supposes data-efficient initialization, as all above-mentioned quantum advantages could be undone when the conversion of classical data to quantum data becomes a bottleneck. 

There are several quantum state preparation algorithms \cite{bergholm2005quantum,Plesch2011, malvetti2021quantum, Araujo0407} with a lower bound of $O(2^n)$ CNOT gates to prepare an arbitrary quantum state with $n$ qubits. 
Attempts to prepare quantum states more efficiently include a divide-and-conquer strategy that exchanges circuit depth by circuit width~\cite{Araujo0407,araujo2021configurable}, probabilistic approaches~\cite{zhang2021low,Park2019}, and strategies to initialize approximated quantum states~\cite{zoufal2019quantum,nakaji2021,javier2021}. 
Most recently, there has been an increasing focus on developing methods that are tailored to specific classes of quantum states. This approach recognizes that not all quantum states are equal, and that different types of states may require different preparation techniques to achieve optimal results. By tailoring the state preparation algorithm to the specific characteristics of the state of interest, these methods can often achieve better fidelity and gate complexity than more general approaches. Examples of this include uniform~\cite{mozafari2021efficient}, sparse~\cite{malvetti2021quantum,mozafari_2022,gleinig_2021,veras2021} and probability distribution~\cite{zoufal2019quantum} states. However, there is no clear understanding of which classes of quantum states can be created efficiently.

Entanglement is a crucial quantum resource that enables the development of more efficient algorithms and protocols. Previous research has demonstrated its relationship to the algorithmic complexity of a quantum state~\cite{MORA2006,PhysRevLett.95.200503}. Entanglement also plays a key role in many quantum applications, such as quantum communication, quantum error correction, and quantum secret sharing~\cite{PhysRevLett.86.5188,Barreiro2010,PhysRevA.54.1098,Steane1996,PhysRevLett.83.648,Cirac2000}. However, the circuit complexity of current algorithms to create a quantum state do not consider the resource of entanglement.

The main goal of this article is to define a deterministic state preparation algorithm that creates circuits with depth as a function of entanglement, and show that for mapping the classical data into a Noisy Intermediate-Scale Quantum (NISQ) device, one indeed wants to be in the regime of low-entanglement. After finalizing the state preparation, one can increase the amount of entanglement (exponential in the number of qubits) by applying unitary evolution. This path could lead to the solution of the so-called loading problem (LP) of classical-quantum information processing.

The proposed approach's advantage is that it can accelerate quantum applications that require the initialization of quantum states, in particular on noisy devices. Using an approximation of the quantum state, the result is that the error introduced by the approximation is smaller than the error to encode the original state. The fundamental cause of this behavior is the difference in the number of noisy operations necessary between the circuits to encode the original versus the approximate state. But also apart from this obvious advantages in the NISQ-era, the proposed approach reduces the complexity to initialize a low-entangled quantum state into fault-tolerant quantum devices.

The remainder of this paper is organized into four sections. The main contributions of this work are described in sections~\ref{sec:entanglement}  and~\ref{sec:baa}.  Section~\ref{sec:entanglement} introduces the \textit{Low-Rank State Preparation} (LRSP) algorithm,
which reduces the computational cost when the input vector has a low Schmidt rank. In this way, the LRSP can be used to initialize general quantum states and also low rank quantum states. Section~\ref{sec:baa} introduces the \textit{Bounded Approximation Error Algorithm} (BAA), a search algorithm that reduces the depth of LRSP circuits given an allowable error. Some numerical tools and the connection to the geometric measure of entanglement are discussed. Section~\ref{sec:lp} presents experimental results on how probability distributions can effectively be created and read out from actual pre-NISQ processors.
Section~\ref{sec:conclusion} summarizes the work and presents perspectives along with potential future research.

\section{Low-Rank State Preparation Algorithm}
\label{sec:entanglement}

\begin{figure*}[ht]
\centering

\begin{subfigure}[b]{.365\textwidth}
    \centering
    \includegraphics[width=.8\textwidth]{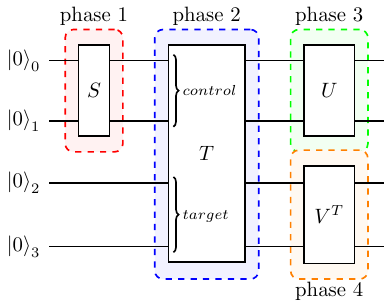}
    \caption{Block diagram}
    \label{fig:plesch_block_diagram}
\end{subfigure}
\begin{subfigure}[b]{.355\textwidth}
    \centering
    \includegraphics[width=.8\textwidth]{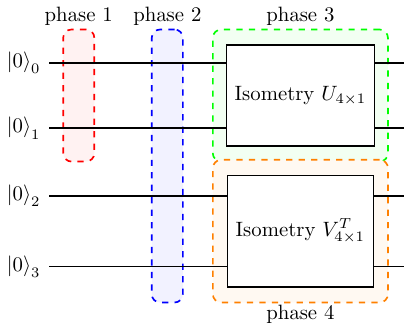}
    \caption{$m=0$}
    \label{fig:plesch_rank_1}
\end{subfigure}

\begin{subfigure}[b]{.400\textwidth}
    \centering
    \includegraphics[width=.8\textwidth]{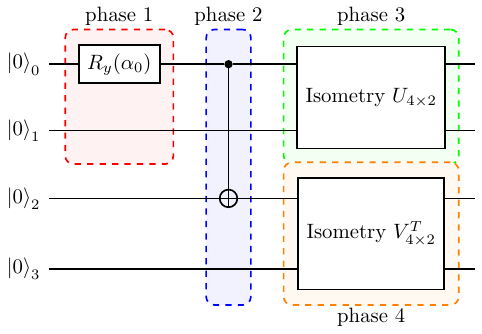}
    \caption{$m=1$}
    \label{fig:plesch_rank_2}
\end{subfigure}
\begin{subfigure}[b]{.500\textwidth}
    \centering
    \includegraphics[width=1.0\textwidth]{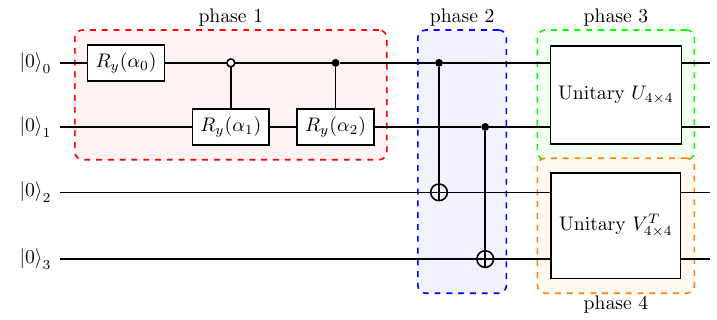}
    \caption{$m=2$}
    \label{fig:plesch_rank_4}
\end{subfigure}
\caption{Schematics of the low-rank approximation algorithm ($n=4$ and subsystem $\mathcal{H}_A=\{0,1\}$). (a) Block diagram circuit. Operator $S$ is responsible for encoding $2^{n_A}$ ($n_A=n/2$) SVD singular values. In this example, operator $S$ encodes a maximum of four amplitudes (two qubits). Operator $T$ is a sequence of CNOTs, controlled by the first half of qubits and targeting the second half (one by one). Operators $U$ and $V$ are the two SVD unitaries or isometries. For full-rank ($m=n_A$, in units of e-bits), $U$ and $V^T$ are encoded as unitaries of dimension $2^{n_A}\times 2^{n_A}$. For a lower rank ($m<n_A$), the operators are encoded as isometries of dimension $2^{n_A}\times 2^m$. When $m=0$ ($\text{rank}=1$), the isometries of dimension $2^{n_A} \times 1$ are equivalent to amplitude encoding with input vectors of length $2^{n_A}$. Plesch's original work describes the four individual phases.
(b)(c)(d) Detailed views of (a) for $\text{rank}=1$ ($m=0$), $\text{rank}=2$ ($m=1$) and $\text{rank}=4$ (full-rank, $m=2$). In this example, phase 1 uses Möttönen's state preparation but could use any amplitude encoding algorithm, including the low-rank state preparation algorithm. }
\label{fig:plesch_low_rank}
\end{figure*}

This work introduces an algorithm called Low-Rank State Preparation (LRSP) algorithm, which is based on the Plesch \& Brukner (PB) state preparation algorithm~\cite{Plesch2011}. The PB algorithm is connected to the Schmidt decomposition, which allows representing a given quantum state $\ket{\psi}$ as a superposition of Schmidt coefficients and corresponding orthonormal basis states in two quantum subsystems, $\mathcal{H}_A$ and $\mathcal{H}_B$. The Schmidt decomposition of $\ket{\psi}$ is expressed as:
\begin{equation}
\label{eq:schmidt_decomposed_state}
    \ket{\psi} = \sum_{i=1}^k \sigma_i \ket{i_A}\ket{i_B}
\end{equation}
Here, $k$ is the Schmidt rank, $\sigma_i$ represents the Schmidt coefficients, ${\ket{i_A}} \in \mathcal{H}_A$ and ${\ket{i_B}} \in \mathcal{H}_B$ are orthonormal bases, and $1 \leq i \leq \min(\dim(\mathcal{H}_A), \dim(\mathcal{H}_B) )$.

The PB algorithm consists of four main steps. The first step involves performing the Schmidt decomposition on a classical computer. In the second step, the algorithm initializes a quantum state in the first register as $\sum_i \sigma_i \ket{i}\ket{0}$, aiming to incorporate the Schmidt coefficients into the state. Following that, in the third step, $\lfloor n/2 \rfloor$ Controlled-NOT (CNOT) gates are applied to create the state $\sum_i \sigma_i \ket{i}\ket{i}$. Finally, in the last step, the algorithm applies the unitary operation $U$ to the first register and the transpose of $V$ (denoted as $V^T$) to the second register, where $U\ket{i} = \ket{i_A}$ and $V^T\ket{i} = \ket{i_B}$.
A quick review of the algorithm is schematically shown in Fig.~\ref{fig:plesch_rank_4}.

The proposed LRSP differs from PB algorithm when the Schmidt measure $m = \lceil \log_2(k)\rceil < \lfloor n/2 \rfloor$ and also by the use of isometries instead of full unitaries.
Theorem~\ref{thm:lrsp} establishes the CNOT gates count needed when a low-rank representation of a state can be found using the Schmidt decomposition, as well as approximating the state by truncating the Schmidt coefficients.

\begin{theorem}[Low-Rank State Preparation]\label{thm:lrsp}
Given Eqn.~(\ref{eq:schmidt_decomposed_state}) with the Schmidt measure $m = \lceil \log_2(k)\rceil$, the low-rank state preparation has a complexity of
\begin{table}[H]
    \centering
    \begin{tabular}{c|c}\hline
        condition & CNOT count \\ \hline
        $0\le m <n_A$ & $O(2^{m+n_B})$ \\ \hline
        $m=n_A$ & $O(2^{n})$ \\ \hline
    \end{tabular}
\end{table}
\end{theorem}
\begin{proof}
When acting on $s$ qubits, a quantum state preparation typically requires $2^s-s-1$ CNOTs~\cite{bergholm2005quantum}, a unitary decomposition $\nicefrac{23}{48}(2^{2s}) - \nicefrac{3}{2}(2^s) + \nicefrac{4}{3}$ \cite{shende2006synthesis}, and an isometry decomposition $2^{m+s}-\nicefrac{1}{24}(2^s)+O(s^2)2^m$ \cite{Iten2016}.
Let $\ket{\psi}$ be a $n$-qubit quantum state with a Schmidt decomposition where subsystem $\mathcal{H}_A$ has $n_A$ qubits ($1\le n_A \le \lfloor n/2 \rfloor$) and subsystem $\mathcal{H}_B$ has $n_B=n-n_A$ qubits. Considering the complete LRSP circuit, the overall number of CNOT gates is represented by:

\begin{footnotesize}
\begin{itemize}[leftmargin=*]

\item $0 \le m < n_A$
\begin{equation*} \label{eq:iso_cnots}
    \underbrace{ 2^m-m-1 \vphantom{\frac{1}{24}} }_\text{phase 1} + \underbrace{ m \vphantom{\frac{1}{24}} }_\text{phase 2} + \underbrace{ 2^{m+n_A}-\frac{1}{24}2^{n_A} }_\text{phase 3 (isometry)} + \underbrace{ 2^{m+n_B}-\frac{1}{24}2^{n_B} }_\text{phase 4 (isometry)}
\end{equation*}

\item $m=n_A$ and $n_A<n_B$
\begin{equation*} \label{eq:uni_iso_cnots}
    \underbrace{ 2^{n_A}-n_A-1 \vphantom{\frac{1}{24}} }_\text{phase 1} + \underbrace{ n_A \vphantom{\frac{1}{24}} }_\text{phase 2} + \underbrace{ \frac{23}{48}2^{2n_A} - \frac{3}{2}2^{n_A} + \frac{4}{3} }_\text{phase 3 (unitary)} + \underbrace{ 2^{n}-\frac{1}{24}2^{n_B} }_\text{phase 4 (isometry)}
\end{equation*}

\item $m=n_A$ and $n_A=n_B$
\begin{equation*} \label{eq:uni_cnots}
    \underbrace{ 2^{n_A}-n_A-1 \vphantom{\frac{1}{24}} }_\text{phase 1} + \underbrace{ n_A \vphantom{\frac{1}{24}} }_\text{phase 2} + \underbrace{2 \left( \frac{23}{48}2^{n} - \frac{3}{2}2^{n_A} + \frac{4}{3} \right) }_\text{phases 3 and 4 (unitaries)}
\end{equation*}
\end{itemize}
\end{footnotesize}
The phases brackets indicate the contribution from each phase of the LRSP procedure to the number of CNOTs. Phase 1 is a state preparation, phase 2 a sequence of CNOT gates, phases 3 and 4 are isometry or unitary decompositions. These equations are bounded by the results of Theorem~\ref{thm:lrsp}.

\end{proof}

With the Schmidt measure $m = \lceil \log_2(k)\rceil < n_A$, the operator $S$ (Fig.~\ref{fig:plesch_block_diagram}) initializes a state with $m$ qubits in phase 1 (Fig.~\ref{fig:plesch_rank_1} and Fig~\ref{fig:plesch_rank_2}), instead of $n_A$ qubits (Fig.~\ref{fig:plesch_rank_4}). 
In phase 2, $m$ CNOT gates are required as they are unnecessary where the control qubit is $\ket{0}$; $m$ quantifies the entanglement between subsystems $\mathcal{H}_A$ and $\mathcal{H}_B$.
Finally, in phases 3 and 4, the matrices $U$ and $V^T$ are isometries $2^m$--to--$2^{n_A}$ and $2^m$--to--$2^{n_B}$, respectively, which require $O(2^{m + n_A})$ and $O(2^{m + n_B})$ CNOTs. The number of CNOTs in the complete LRSP circuit is $O(2^{m + n_B})$ because the cost of isometry $V^T$ (phase 4) dominates the cost of the algorithm ($n_B \ge n_A$).

In the best-case scenario, when the bipartition is not entangled, $\ket{\psi} = \ket{\psi_A}\ket{\psi_B}$, the rank is equal to 1 ($m=0$ e-bits), and phase 1, which encodes singular values, can be skipped entirely. Phase 2 requires no entanglement between the subsystems, as there are zero e-bits between them. Phases 3 and 4 involve one $1$--to--$2^{n_A}$ and one $1$--to--$2^{n_B}$ isometry, respectively, as shown in Fig.\ref{fig:plesch_rank_1}. These isometries are equivalent to two parallel sub-state preparations in $n_A$ and $n_B$ qubits. If the original state is a product state, applying the same algorithm recursively to prepare the sub-states in phases 3 and 4 generates a circuit without CNOT gates. If the original state is not completely separable, the cost of the state preparation is $O(2^{n_e})$, where $n_e$ is the number of qubits in the most entangled subsystem.

In the worst-case scenario, where $m=n_A$, the LRSP algorithm generates circuits with a depth of $O(2^n)$, and is competitive with other state-of-the-art deterministic algorithms for general state preparation without auxiliary qubits. 

Table~\ref{tab:sp_comparison-worstcase} compares the LRSP circuit depth and number of CNOTs with previous state preparation algorithms for both separable and non-separable quantum states. The selection of states is done randomly while ensuring that they satisfy the criteria of being non-separable and completely separable. It is important to note that the results will remain consistent for any state with similar entanglement characteristics. The specific counts of the LRSP method are influenced by the choice of unitary and isometry decompositions. The methods used to produce Table~\ref{tab:sp_comparison-worstcase} include Quantum Shannon Decomposition (QSD) for unitaries~\cite{shende2006synthesis}, Cosine--Sine Decomposition (CSD)~\cite{Iten2016} for $2^{n-1}$--to--$2^n$ isometries, and Column--by--Column Decomposition (CCD) for other isometries~\cite{Iten2016}.

\begin{table}[]
    \resizebox{\columnwidth}{!}{
    \centering
    \begin{tabular}{c|c|c|c|c|c|c}
        \hline        
         $n$ & Method                     &  Script                   & \multicolumn{2}{c|}{Entangled}    & \multicolumn{2}{c}{Separable}     \\
             &                            &                           & \multicolumn{1}{c}{CNOTs} & Depth & \multicolumn{1}{c}{CNOTs} & Depth \\ \hline
         \multirow{4}{*}{10} & Low Rank                   &     \cite{qclib}          & 915 & 1047 & 0 & 2 \\
         & PB \cite{Plesch2011}       & \cite{qclib}              & 919 & 1057 & 919 & 1058   \\
         & Isometry \cite{Iten2016}   &  \cite{Qiskit}            & 1013 & 4055 & 1013 & 3871 \\
         & Multiplexor \cite{shende2006synthesis} &  \cite{Qiskit}& 2026 & 4064 & 2026 & 4064 \\ \hline
         \multirow{4}{*}{15} & Low Rank                   &     \cite{qclib}          & 30998 & 53644 & 0 & 2 \\
         & PB       & \cite{qclib}              & 38814 & 71580 & 38811 & 71575   \\
         & Isometry   &  \cite{Qiskit}            & 32752 & 65505 & 32752 & 65505   \\
         & Multiplexor &  \cite{Qiskit}& 65504 & 131025 & 65504 & 131025 \\ \hline
    \end{tabular}
    }
    \caption{Depth and number of CNOTs comparison between LRSP and other state preparation algorithms. The column "Entangled" indicates that the state is non-separable, and the column "Separable" indicates that the state is completely separable (a product state). These numerical experiments were performed without Qiskit's circuit optimization.}
    \label{tab:sp_comparison-worstcase}
\end{table}

\subsection{Low-Rank Approximation}
The LRSP algorithm also allows a low-rank approximation limiting the Schmidt rank in exchange for an error. The fidelity loss can be used to quantify the loss by the approximation. 
\begin{definition}
Given the low-rank parameter $r$, the approximated state is denoted as 
\begin{equation}
\label{eq:schmidt_decomposted_approximated_state}
    \ket{\psi^{(r)}} = \sum_{i=1}^r \sigma_i \ket{i_A}\ket{i_B}
\end{equation}
with coefficients for $1 \leq r \leq k$, i.e. $\sigma_j=0$, $r < j \leq k$. 
\end{definition}

\begin{figure}[ht]
    \centering
    \includegraphics[width=0.75\columnwidth]{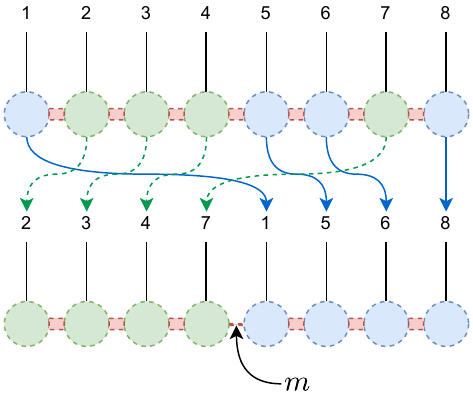}
    \caption{Schematic of the logical swap of qubits. By rearranging the qubits it is possible to find a bipartition such that the bond dimension between the partition blocks is small, which means that these blocks are not strongly entangled. The indices indicate the position of the qubit in the quantum circuit, and the color (blue or green) indicate the bipartition block.}
    \label{fig:rearranging_sites}
\end{figure}

It is possible to partially ($r>1$) or completely ($r=1$) disentangle subsets of qubits while the introduced fidelity loss $l(r, \ket{\psi}) := (1 - \vert\langle \psi, \psi^{(r)} \rangle\vert^2) = \sum_{i=r+1}^{k} \vert\sigma_i\vert^2$ scales with the Schmidt coefficients that are dropped. The remaining coefficients must be normalized.

Finding the optimal configuration for disentangling arbitrary bipartition blocks is not immediately apparent. However, by rearranging qubits between blocks of any size within the allowed range ($1\le n_A \le \lfloor n/2 \rfloor$ and $n_B=n-n_A$), it is possible to identify the configuration that exhibits a low bond-dimension, enabling a lower-error approximation and potentially disentangling these blocks (Fig.~\ref{fig:rearranging_sites}). This approach can be recursively applied on the two resulting blocks separately, and a search algorithm with a given maximal approximation error can then find the optimal approximation of any quantum state of interest, including, for example, vector encoding for inverting matrices using HHL~\cite{lloyd2014quantum}, loading data into a quantum machine learning model~\cite{biamonte2017quantum, schuld2018supervised} or using quantum simulation of stochastic processes~\cite{blank2021quantum}. The resulting search algorithm is called the Bounded Approximation error Algorithm (BAA).

\section{Bounded Approximation Algorithm}
\label{sec:baa}

By design, the low-rank approximation only applies to bipartite systems, yet it can be used hierarchically to enable the analysis of multipartite quantum systems~\cite{10.1007/s11128-017-1633-8} by a recursive algorithm. One then can recursively apply this approach on the two resulting partitions separately, i.e., find in the best rearranging of sites to minimize a bond dimension. This method leads to a tree search algorithm. With a given maximal approximation error, one can then find the optimal approximation of a quantum state. It is a bounded approximation error state preparation algorithm (BAA) that has a classical exponential run-time with respect to the number of qubits of the state as an upper bound. As it is a branch-and-bound algorithm using breadth-first search, the complexity usually converges faster~\cite{mehlhorn_algorithms_2008,skiena_algorithm_2008}, especially when the maximum allowed error is low ($\sim 0.01$) and the algorithm terminates at the first levels of the search tree. The full set of pseudocode which describes the algorithm is printed in the supplementary information. In what follows, we want to outline the core principles of the algorithm.

\begin{figure*}[ht]
    \centering
    \includegraphics[width=.95\textwidth]{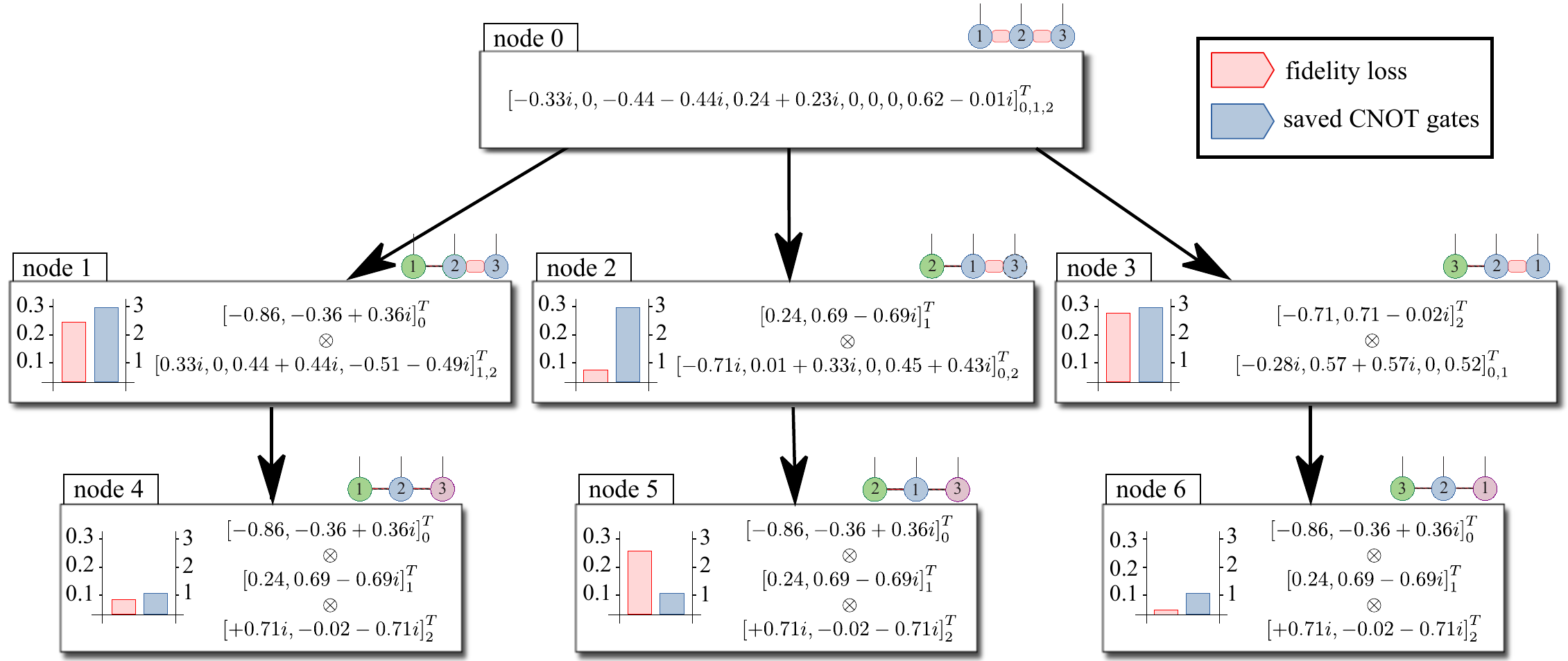}
    \caption{An example of a three-qubit state that is disentangled into product states (leaves). It starts with one vector in the root and then all $\mathcal{B}(3)=3$ bipartitions are branched off, creating each two states. From there, the larger subspace has two qubits, so it should have $\mathcal{B}(2)=2$ branches. But in this case, the second branch is equal to the complement of the first, so it is redundant and omitted. Each node has a fidelity loss (red column) and the number of CNOT gates saved (blue column) using the proposed algorithm. The sub-indices of the vectors are the qubits associated with each state. Adding the fidelity losses from root to any leaf gives the total fidelity loss of $\approx0.307$, meaning that the reduced product state to its original state have an overlap of $\approx 0.693$, but with a saving of 4 CNOT gates. The adaptive approximation algorithm goes to ask if we can save some CNOT gates with an acceptable fidelity loss. If $l_\text{max}=0.1$, there is the second bipartition from the root node (node 2) that introduces only a fidelity loss of $\approx 0.058$ by simultaneously saving three CNOT gates.
    }
    \label{fig:tree_fidelity_loss}
\end{figure*}

\begin{figure*}[ht]
    \centering
    \includegraphics[width=.8\textwidth]{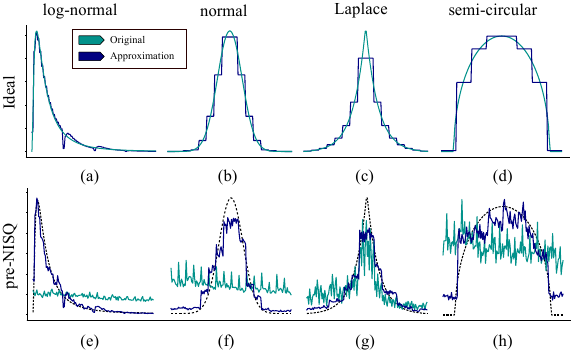}
    \caption{Visualization of probability distributions encoded on the amplitudes of a 7-qubit quantum state. (a)-(d) Ideal values of the exact (green line) and approximate (blue line) distributions. (e)-(h) Values of the exact and approximate distributions encoded via BAA and estimated by measurements on ibm\_perth device. In the actual device, the encoding of the approximations performs closer to the ideal (black dotted line) than the exact distribution encoding. Each result is an average of 10 runs with 8192 shots.
    }
    \label{fig:distributions}
\end{figure*}

\begin{figure*}[ht]
    \centering
    \includegraphics[width=.9\textwidth]{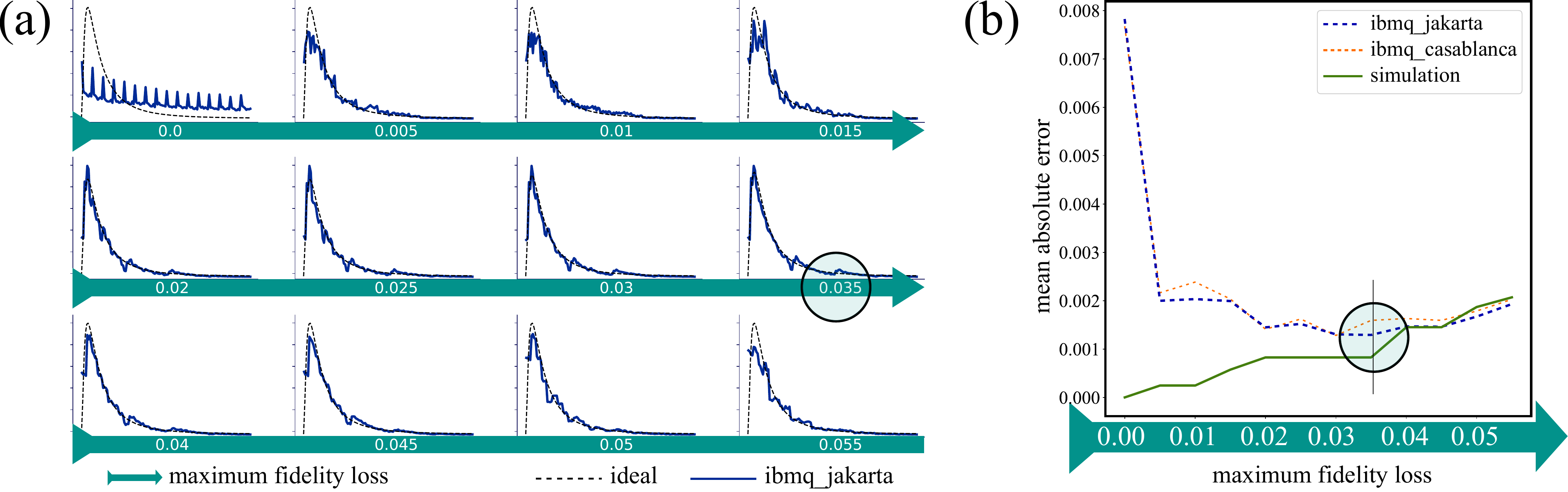}
    \caption{BAA experimental results with a 7-qubit log-normal state on two IBM quantum devices (ibmq\_jakarta \& ibmq\_casablanca), with a geometric entanglement of $\approx0.0534$. (a) Several approximations in ascending order show that the exact state preparation fails. All subsequent approximations decrease the MAE until $l_{max}=0.035$. After that, it increases until the product-state (at $l_{max}=0.055$) is reached. (b) Comparison between two devices and an ideal simulation. The highlighted point indicates the lowest MAE achieved by the devices.
    }
    \label{fig:initialization_experiments}
\end{figure*}

The approximate state preparation algorithm works as follows. Any (pure) $n$-qubit quantum state $\ket{\psi}$ will allow for a total of
\begin{equation} \label{eq:comb}
    \mathcal{B}(n) = \sum_{t=1}^{\lfloor n/2\rfloor} \left(
    \begin{matrix} n \\ t
    \end{matrix}\right)
\end{equation} 
bipartitions (also called branches), each is written as $b \in \left[1 \isep \mathcal{B}(n) \right]$ with the Schmidt number bounded by $k(b) = \min\{\dim(\mathcal{H}_1^b), \dim(\mathcal{H}_2^b)\}$. By letting the low-rank parameter $1 \leq r \leq k(b)$ be $r = 1$, each bipartition creates a disentangled pair of smaller states. Combining this pair by the tensor-product returns an approximation of the original state. 
Additionally, it is possible by the complexity analysis of Theorem~\ref{thm:lrsp} to predict the number of CNOT gates saved.

Starting from $\ket{\psi} \in \mathcal{H}$, the algorithm is branching into all possible bipartitions, $(\mathcal{H}_1^b, \mathcal{H}_2^b)$. Recursively, each partition $\mathcal{H}_i^b$ can in turn be branched once again, as long as $\mathcal{H}_i^b$ ($i=1,2$) is not one single qubit or the total fidelity loss has not exceeded the error bound $l_\text{max}$. In order for the search-algorithm to work, each node needs to be able to know the total fidelity loss, the total saved number of CNOT gates and its partitioned Hilbert spaces. Due to this structure, the algorithm terminates in the worst case with an exponential number of steps. These mentioned properties are summarized.
\begin{lemma}\label{lemma:recursive_fidelity}
The fidelity and the saved CNOT gates of each branch can be recursively calculated.
\end{lemma}
A small result with an impact is the fact that the fidelity/fidelity loss and CNOT gate savings can be recursively calculated from one branch to another. The lemma is important for the computation of the search tree in the BAA algorithm, as it is possible to parallelize it.
\begin{proof}
One must show that the total fidelity loss over several steps in the tree can be recursively calculated. Let $l_{b_p}$ and $l_{b_c}$ be the fidelity loss of the parent and child node, respectively. Then the fidelity of both is $(1 - l_{b_p})(1 - l_{b_c})$, hence $l_\text{tot} = 1 - (1 - l_{b_p})(1 - l_{b_c})$. More generally, $l_\text{tot} = 1 - \prod_i (1-l_{b_i})$ for a path on the tree $b_1, \ldots, b_t$.
\end{proof}

\begin{proposition}\label{prop:baa_algorithm_complexity}
The BAA-Algorithm's search tree has $n$ levels, and therefore terminates. The time and space complexity is exponential in $n$ in its worst case.
\end{proposition}
When spanning the search tree (see Suppl. Information Pseudocode 1), each child of a parent node will be created by choosing exactly one state and doing exactly one bipartition. The bipartitions of our focus are the $t$ qubits vs. the rest ones, where $t$ is defined in Eq.~\eqref{eq:comb}. Say, among the $k$ states of the parent node, $\# 1$ is selected to be bipartitioned, with $\dim(\mathcal{H}_1) = n_1$. The split of $t$ qubits vs. rest is done, and in this manner $n_1 - 1$ levels will be traversed until this state has been partitioned to a tensor product of single qubit states. Meanwhile, all $k-1$ original states were kept as they were. Now the partition $\# 2$ is selected, and in turn takes $n_2 - 1$ steps to a tensor product of single qubit states. This goes on with all the remaining $k-2$ states. So in total, there are 
\begin{equation}
    \sum_{l=1}^k (n_l - 1) = n - k
\end{equation}
steps until all $k$ sates have been partitioned into tensor products of single qubit Hilbert spaces. As this is the longest path possible, the BAA algorithm terminates. Now, we show the runtime complexity of the BAA algorithm.
\begin{lemma}
Given a $k$-fold partitioning of $\mathcal{H}$, denotes as $(\mathcal{H}_1, \ldots, \mathcal{H}_k)$, the number of children of this node is 
\begin{equation}
\begin{aligned}
    \mathcal{B}_k(n) &= \sum_{i=1}^{k} \mathcal{B}(\dim{\mathcal{H}_i})
\end{aligned}
\end{equation}
As a special case, we see $k=1$, which is directly $\mathcal{B}(\dim{\mathcal{H}})$.
\end{lemma}
\begin{proof}
Given $k$ already partitioned subspaces, the search algorithm selects one of the parts ($\mathcal{H}_i$, $i=1, \ldots, k$) and carries out all $\mathcal{B}(\dim{\mathcal{H}_i})$ branches, while leaving all the other $k-1$ original parts intact. This is then repeated, so the total number of children this node has as claimed.
\end{proof}
After each step, the number of parts thus changes from $k \mapsto k + 1$ as one part is created. Of course, the sum of all dimensions must be equal to the dimension of the original Hilbert space $n$. Thus, each level of the search tree is identified by the number $k$ of parts. The maximum number of levels is therefore always $n$. The total number of nodes is then computed as follows.

The first level has exactly $1$ node. The second level has $\mathcal{B}_1(n)$ nodes. The third level has $\sum_{b=1 \ldots \mathcal{B}_1(n)} \mathcal{B}_2(n; b)$ and the fourth level $\sum_{b_1=1 \ldots \mathcal{B}_1(n)} \sum_{b_2=1 \ldots \mathcal{B}_2(n; b_1)} \mathcal{B}_3(n; b_2)$ up until the $n^\text{th}$ level, we have,
\begin{equation}
    \sum_{b_1=1}^{\mathcal{B}_1(n)} \cdots \sum_{b_{n-1} = 1}^{\mathcal{B}_{n-1}(n; b_{n-2})} \mathcal{B}_n(n; b_{n-1})
\end{equation}
This is clearly exponential in $n$. A tree of an example three-qubit state is shown in Fig.~\ref{fig:tree_fidelity_loss}. With an $l_\text{max}=0.1$ only one bipartition is possible within this error bound, and three CNOT gates are saved.

Albeit the fact that the BAA algorithm uses low-rank approximations with $r=1$ to fully disentangle the states (Fig.~\ref{fig:plesch_low_rank} and Fig.~\ref{fig:tree_fidelity_loss}), it is possible to include low-rank approximations with $r>1$ when complete disentanglement is no longer within $l_\text{max}$. This allows fine-tuning the partial disentanglement of states, as each value of the parameter $r$ may achieve reductions in the number of CNOTs that would not be possible with the original approach, producing additional branches from the tree nodes. Therefore, this advantage comes with a larger search space, and increases the cost of the algorithm.

While the proposed approximate state preparation algorithm reduces the quantum circuit complexity, the BAA has an exponential preprocessing cost. The naive (``brute-force'' breadth-first) algorithm is already a solution for those researchers who need to work on small-qubit experiments and currently have no alternatives to an efficient quantum initialization on noisy devices. But a scalable workaround is necessary. One way is to use a greedy approach~\cite{Nemhauser1978,cormen_introduction_2009} that seeks to reduce the computational cost, making BAA a useful solution for general problems.

The greedy strategy proposes that branching from a node is limited to a qubit-by-qubit analysis, selecting only one representative of the blocks of size k where $1 \le k \le \lfloor n/2 \rfloor$ instead of the full combination as shown in Equation~\eqref{eq:comb}. The increment in the block size is done by choosing the qubit with the lowest fidelity loss when removed from the remaining entangled subsystem. As an example, on a seven qubit state, the best 1 vs. 6, 2 vs. 5 and 3 vs. 4 bipartitions are attempted. Among these, only the best one is selected locally and then propagated in the recursion, producing a single path. This approach generates a tree with a linear number of nodes on $n$.
The greedy algorithm reduces the search-problem exponentially and will never exceed the approximation configured, but the downside is that the CNOT complexity to achieve this approximation may not be optimal.

\subsection{Product State Approximations}

In view of treating quantum state approximation, one could be naturally inclined to ask for the best product state approximation. This is indeed a measure for entanglement, the so-called geometric measure of entanglement. It was first introduced by Shimony~\cite{10.1111/j.1749-6632.1995.tb39008.x} and refined by Barnum \& Linden~\cite{10.1088/0305-4470/34/35/305}. The definition for this measure of entanglement is given by
\begin{equation}
    E_g(\ket{\psi}) = \min_{\ket{\phi}} \| \ket{\psi} -\ket{\phi} \|
\end{equation}
where the minimization is over all states $\ket{\phi}$ that are product states~\cite{10.1007/s11128-017-1633-8}, i.e., $\ket{\phi} = \otimes_{l=1}^n \ket{\phi^l}$ with each $\ket{\phi^l}$ being a local state. That this value is a useful measure of entanglement was shown by Wei et al.~\cite{10.1103/physreva.68.042307}, it is an entanglement monotone.

The geometric measure of entanglement is related to the bounded approximate algorithm (BAA) in the following way. Theoretically, the greatest fidelity loss that any quantum state can experience under the BAA should never be higher than it. Therefore, the geometric entanglement indicates how far an approximation can get. This shows that low-entanglement states can be encoded with low loss as a product state.

\section{The Loading Problem on NISQ Devices}
\label{sec:lp}

The LRSP algorithm is designed to solve a class of problems on NISQ devices, namely loading data. To solve (or avoid) the loading problem~\cite{biamonte2017quantum}, an approximation of the quantum state that uses less entanglement is created. In this section, we present an application of the LRSP to initialize Probability Distribution Functions. In an experimental evaluation, the approximated circuit created with the LRSP algorithm allows for improvements in the presence of noise.

\subsection{Experiments on Probability Density Functions}

The BAA proves to be efficient when working with specific classes of quantum states found in the quantum finance~\cite{malvetti2021quantum,zoufal2019quantum,mozafari2021efficient}. Indeed, encoding probability distributions on the amplitudes of a 7-qubit quantum state show that even a modest fidelity loss ($\le 0.02$) results in a significant reduction in the number of CNOTs.

To encode a continuous probability density function (PDF) into the amplitudes of a quantum state, it is necessary to construct its discrete analog. In general, this construction is based on preserving one or more properties of the continuous distribution~\cite{Chakraborty2015}.

There are several methods by which a discrete random variable can be constructed from a continuous one. Here we employ Methodology-V described in Ref.~\cite{Chakraborty2015} where the cumulative distribution function (CDF) of a discrete random variable $Y$ maintains the form of the CDF of a continuous random variable $X$. The probability mass function (PMF) of $Y$ is built from the CDF of $X$ $F_X(x)=Pr(X\le x)$ and is given by
\begin{equation} \label{eq:pmf}
    P(Y=k)=F_X(k+\delta)-F_X(k-[1-\delta])
\end{equation}
where $0<\delta<1$ and $k=\{0,1,2,\dots\}$.

In the experiments, the interval $0\le x \le 20$ is divided into $2^7$ discretization points. Therefore, the distance between consecutive points is $d=20/(2^7-1)$. By choosing $\delta=d/2$, the parameters of Eq.~\eqref{eq:pmf} are set to
\begin{equation} \label{eq:pmf2}
    P(Y=k)=F_X(k+d/2)-F_X(k-d/2)
\end{equation}
for $k=\{0, d, 2d, \dots, (2^7-1)d\}$.

Equation~\eqref{eq:pmf2} is used to generate the discrete probability distributions analogous to the continuous ones characterized by the normal, log-normal, Laplace and semicircular PDFs.
\begin{table}[ht]
    \centering
    \begin{tabular}{c|c|c|c|c}\hline
        distribution & mean & variance & skewness & kurtosis \\ \hline
        normal & 10.0 & 4.0 & 0.0 & 0.0 \\ \hline
        log-normal & 3.297 & 18.68 & 6.185 & 110.9 \\ \hline
        Laplace & 10.0 & 8.0 & 0.0 & 3.0 \\ \hline
        semicircular & 10.0 & 16.0 & 0.0 & -1.0 \\ \hline
    \end{tabular}
    \caption{First four moments of the probability distributions.}
\end{table}

Each point of the discretization corresponds to an amplitude of the state vector. Note that, given the conservation of probability, each amplitude must be the square root of the PMF at these points.
The result of the discretization is employed to constructs Figure~\ref{fig:distributions} and Figure~\ref{fig:initialization_experiments}.

The state vectors of the distributions were encoded via BAA and estimated by measurements on actual quantum devices. All experiments make use of the breadth-first search (brute force) strategy. Each result is an average of 10 runs with 8192 shots each.
The experiments leading to Figure~\ref{fig:distributions} were performed at points $l_\text{max}=\{0.0, 0.02\}$ using device ibmq\_perth, while those leading to Figure~\ref{fig:initialization_experiments} were performed at points $l_\text{max}=\{0.0,0.005,0.01,\dots,0.055\}$ using devices ibmq\_jakarta and ibmq\_casablanca. All experiments, with and without approximation, were run with Qiskit's standard circuit optimization.

When preparing the original states ($l_\text{max}=0.0$), the distributions need circuits with 105, 109, 28, and 99 CNOTs, respectively. The difference between the CNOT numbers is due to the LRSP's ability to adjust the complexity of the circuit to the degree of state entanglement. If a maximum fidelity loss of $l_\text{max}=0.02$ is allowed, the effective fidelity loss achieved by the BAA to encode the approximate state for each distribution is the closest possible to $l_\text{max}$, i.e., 0.0089, 0.0159, 0.0081, and 0.0158. With such approximation, the BAA can encode the first three distributions using only \textit{six} CNOTs and the semicircular with \textit{three} CNOTs. This relatively low introduced error compared to the extreme reduction of entangling operations leads to better results in terms of mean-absolute-error (MAE). Indeed, when executing an experiment on the ibm\_perth, the high number of CNOTs to initialize the exact state obscure the probabilities whereas the BAA approximation keep distribution specific features closer to the actual distribution, see Figures~\ref{fig:distributions}(a)--\ref{fig:distributions}(h). 

\begin{figure*}[ht]
    \centering
    \begin{subfigure}[b]{1.0\textwidth}
        \centering
        \includegraphics[width=0.45\textwidth]{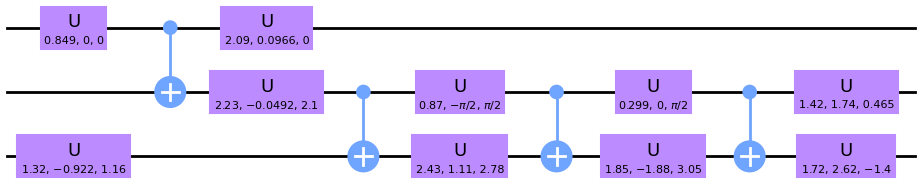}
        \caption{Low-rank}
        \label{fig:lowrank_circuit}
    \end{subfigure}
    \begin{subfigure}[b]{1.0\textwidth}
        \centering
        \includegraphics[width=1.0\textwidth]{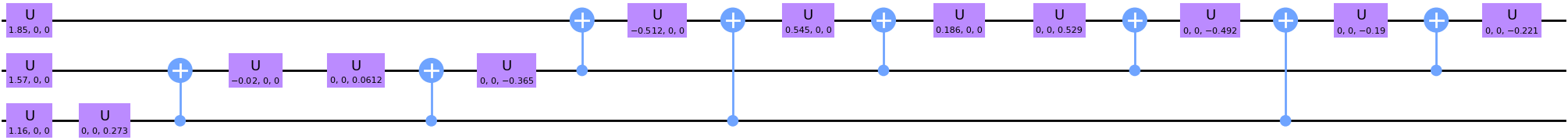}
        \caption{Qiskit}
        \label{fig:qiskit_circuit}
    \end{subfigure}
    
    \caption{Circuits for eight-dimensional exact quantum state initialization. (a) Circuit produced by the LRSP method. (b) Circuit produced by Qiskit's standard initialization.}
    \label{fig:tomography_circuits}
\end{figure*}

To obtain these results, the brute-force algorithm explores 44, 45, 44, and 394 nodes in the search tree, respectively. In contrast, the greedy algorithm achieves the exact same results for these specific distributions but traverses only 4, 4, 4, and 5 nodes, significantly reducing the cost compared to the brute-force strategy.

Given this context, the low-rank state preparation method with the BAA search provides an efficient and competitive method to load a PDF into a noisy quantum device.

\subsection{State Tomography}

The efficiency of the Low-Rank State Preparation (LRSP) method can be assessed through quantum state tomography. For this experiment, we randomly selected the following complex state vector:
\begin{align*}
(0.1619+0.2599i, 0.4111+0.3061i, 0.3165+0.0089i, \\
0.2588+0.4194i, 0.0675+0.3599i, 0.0674+0.0918i, \\
0.0251+0.0786i, 0.3745+0.0793i)
\end{align*}

To compare the LRSP result with Qiskit state preparation under noisy simulation, we analyze the circuits generated by both methods, as depicted in Figure~\ref{fig:tomography_circuits}. Due to the lower number of CNOT gates and reduced depth, LRSP is expected to outperform the traditional amplitude encoding approach, achieving higher fidelity in the presence of noise.

\begin{figure}[ht]
\centering

\begin{subfigure}[b]{1.0\columnwidth}
    \centering
    \includegraphics[width=1.0\textwidth]{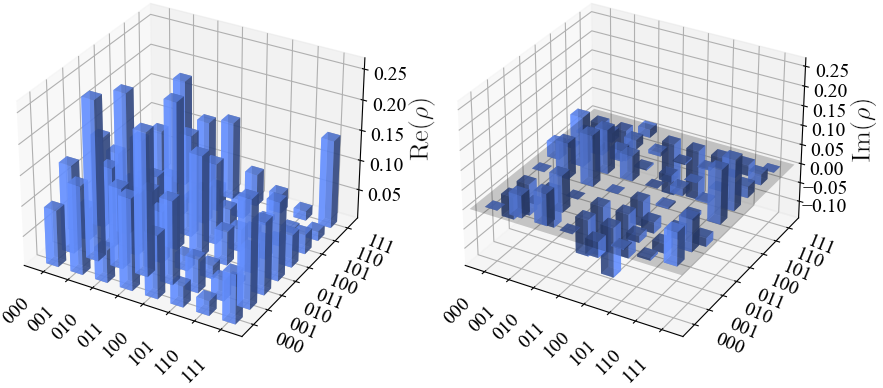}
    \caption{Ideal}
    \label{fig:ideal_dm}
\end{subfigure}
\begin{subfigure}[b]{1.0\columnwidth}
    \centering
    \includegraphics[width=1.0\textwidth]{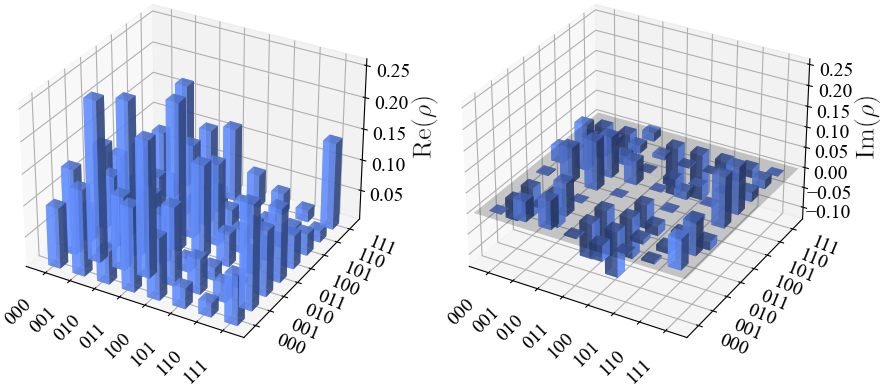}
    \caption{Low-rank}
    \label{fig:lowrank_dm}
\end{subfigure}
\begin{subfigure}[b]{1.0\columnwidth}
    \centering
    \includegraphics[width=1.0\textwidth]{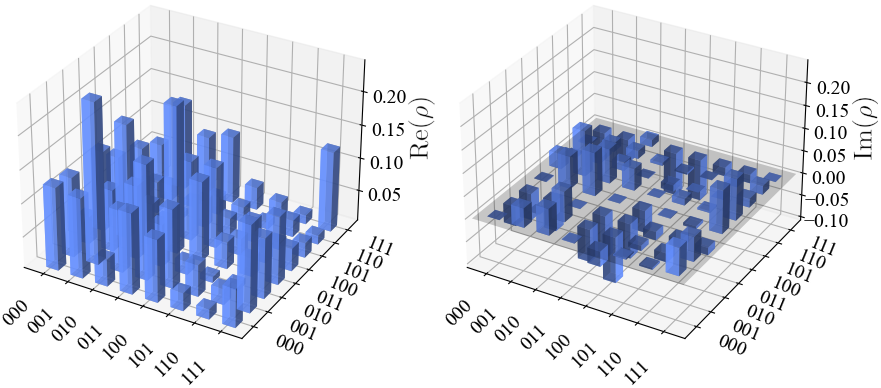}
    \caption{Qiskit}
    \label{fig:qiskit_dm}
\end{subfigure}

\caption{State tomography experiment. (a) Ideal density matrix, produced directly from state vector. (b) Density matrix produced by quantum tomography using noisy simulation and LRSP to prepare the state. (c) Density matrix produced by quantum tomography using noisy simulation and Qiskit's standard state preparation.}
\label{fig:tomography}
\end{figure}

We can visualize in Figure~\ref{fig:tomography} that the state prepared by LRSP closely approximates the ideal result, highlighting the advantage of the proposed method even for small states with only three qubits. It showcases superior performance compared to traditional amplitude encoding methods. The better performance becomes clear when comparing the fidelity of the tomography, 0.92 for LRSP versus 0.78 for Qiskit.

To emulate the behavior of a real device, we utilized Qiskit's ibm\_cairo system snapshot (FakeCairo) in the noisy simulation. A total of 8192 shots were employed for each state initialization method.

\section{Conclusion}
\label{sec:conclusion}

The LRSP algorithm achieves significant reductions in the depth and number of CNOT gates required for quantum state initialization by leveraging the entanglement structure of the quantum state during state preparation. This work highlights that classical data can be rearranged such that the entanglement structure attains an easier to approximate structure, as can be seen in Figure~\ref{fig:rearranging_sites} and Figure~\ref{fig:tree_fidelity_loss}. This directly bestows the data with the topological structure of qubit systems. In particular, by logically swapping qubits between partition blocks, one can find more local behavior. This makes it easier to simulate classically, but also, by virtue of the above-mentioned circuit designs, less complex to create on a quantum computer. To showcase the immediate advantages of this approach, we present an example application in loading PDFs.

Recent approaches for approximating quantum states use variational quantum circuits in which parameterized single qubit and entangling operations are intertwined and adjusted according to a gradient descent method while minimizing/maximizing a loss function, usually taken from the measurement output of an experiment. In particular, qGAN~\cite{zoufal2019quantum} and AAS~\cite{nakaji2021} are noteworthy methods proposed as viable alternatives to standard state preparation, aimed at achieving promising results in the NISQ era. These methods try to emulate local-interaction Hamiltonians, and as such they in theory must be able to converge to the desired quantum state.
However, unlike those algorithms, the BAA is based on purely classical algorithms being very efficient in today's hardware with standard methods to apply for approximations (instead of a full search, we have greedy and Monte-Carlo as options) and the method does not suffer from initial conditions failures (such as barren plateaus).

As a future work, sparse states, known for their low entanglement, can also benefit from the use of LRSP and BAA for state preparation. Introducing sparse isometry and unitary decompositions can further enhance these algorithms~\cite{malvetti2021quantum}, leading to specialized sparse versions. Although BAA already demonstrates potential for preparing sparse states even without these improvements, it is not competitive when compared to specialized methods for states with few non-zero amplitudes~\cite{mozafari_2022,gleinig_2021,veras2021}.
The main advantage of a sparse version would be the capability to process sparse input vectors, where storage space is proportional to the non-zero amplitudes. In contrast, the method proposed in this paper is dense, resulting in storage space always being proportional to $2^n$, even if a portion of the amplitudes are zero. By employing sparse decomposition techniques, the memory and time complexity requirements for running the singular value decomposition (SVD) can also be significantly reduced. In contrast, the use of dense decomposition routines demand $\sim 2^n$ bytes of memory and $O(2^{3n/2})$ time complexity~\cite{KOGBETLIANTZ_1955,forsythe_1960} for $n$-qubit states. These limitations may hinder the preparation of states by the proposed method when $n$ is large.

This work also leaves some open technical questions. For instance, the BAA algorithm needs to disentangle the first level of the search tree to proceed to the next level. However, for some quantum states, finding a bipartition that breaks the system into two uncorrelated parts and results in a fidelity loss less than the maximum allowed may not be possible. Although the Low-Rank state preparation can reduce gate complexity, it may not fully exploit the quantum state's entanglement topology.

In addition to its advantages for state preparation in the NISQ era and beyond, this approach also shows promise for enhancing quantum adiabatic state preparation~\cite{10.1103/revmodphys.90.015002,10.1063/1.4880755}. By preparing a low-entanglement bounded approximation error state and finding its Hamiltonian for which it is an eigenvalue, one could potentially skip spectral gap bottlenecks and reduce the adiabatic time significantly. This approach could even lead to high-entanglement states, although the specific method for achieving this remains unclear. Nonetheless, the potential benefits of this application are significant.

\section*{Acknowledgments}

This research is supported by Conselho Nacional de Desenvolvimento Científico e Tecnológico - CNPq, Coordenação de Aperfeiçoamento de Pessoal de Nível Superior (CAPES), Fundação de Amparo à Ciência e Tecnologia do Estado de Pernambuco - FACEPE, and National Research Foundation of Korea (Grant No. 2022M3E4A1074591). We acknowledge the use of IBM Quantum services for this work. The views expressed are those of the authors, and do not reflect the official policy or position of IBM or the IBM Quantum team.

\section*{Data availability}

The sites \url{https://github.com/qclib/qclib-papers} and \url{https://github.com/qclib/qclib} contain all the data and the software generated during the current study.

\bibliographystyle{IEEEtran}
\bibliography{references}

\begin{thebibliography}{10}
\providecommand{\url}[1]{#1}
\csname url@samestyle\endcsname
\providecommand{\newblock}{\relax}
\providecommand{\bibinfo}[2]{#2}
\providecommand{\BIBentrySTDinterwordspacing}{\spaceskip=0pt\relax}
\providecommand{\BIBentryALTinterwordstretchfactor}{4}
\providecommand{\BIBentryALTinterwordspacing}{\spaceskip=\fontdimen2\font plus
\BIBentryALTinterwordstretchfactor\fontdimen3\font minus
  \fontdimen4\font\relax}
\providecommand{\BIBforeignlanguage}[2]{{%
\expandafter\ifx\csname l@#1\endcsname\relax
\typeout{** WARNING: IEEEtran.bst: No hyphenation pattern has been}%
\typeout{** loaded for the language `#1'. Using the pattern for}%
\typeout{** the default language instead.}%
\else
\language=\csname l@#1\endcsname
\fi
#2}}
\providecommand{\BIBdecl}{\relax}
\BIBdecl

\bibitem{arute2019quantum}
F.~Arute, K.~Arya, R.~Babbush, D.~Bacon, J.~C. Bardin, R.~Barends, R.~Biswas,
  S.~Boixo, F.~G. Brandao, D.~A. Buell \emph{et~al.}, ``Quantum supremacy using
  a programmable superconducting processor,'' \emph{Nature}, vol. 574, no.
  7779, pp. 505--510, 2019.

\bibitem{harrow2009quantum}
A.~W. Harrow, A.~Hassidim, and S.~Lloyd, ``Quantum algorithm for linear systems
  of equations,'' \emph{Physical review letters}, vol. 103, no.~15, p. 150502,
  2009.

\bibitem{lloyd2014quantum}
S.~Lloyd, M.~Mohseni, and P.~Rebentrost, ``Quantum principal component
  analysis,'' \emph{Nature Physics}, vol.~10, no.~9, pp. 631--633, 2014.

\bibitem{10.1098/rspa.2015.0301}
A.~Montanaro, ``{Quantum speedup of Monte Carlo methods},'' \emph{Proceedings
  of the Royal Society A: Mathematical, Physical and Engineering Sciences},
  vol. 471, no. 2181, p. 20150301, 2015.

\bibitem{10.1103/physreva.98.022321}
P.~Rebentrost, B.~Gupt, and T.~R. Bromley, ``{Quantum computational finance:
  Monte Carlo pricing of financial derivatives},'' \emph{Physical Review A},
  vol.~98, no.~2, p. 022321, 2018.

\bibitem{biamonte2017quantum}
J.~Biamonte, P.~Wittek, N.~Pancotti, P.~Rebentrost, N.~Wiebe, and S.~Lloyd,
  ``Quantum machine learning,'' \emph{Nature}, vol. 549, no. 7671, p. 195,
  2017.

\bibitem{schuld2018supervised}
M.~Schuld and F.~Petruccione, \emph{Supervised Learning with Quantum
  Computers}, 1st~ed.\hskip 1em plus 0.5em minus 0.4em\relax Springer
  Publishing Company, Incorporated, 2018.

\bibitem{10.1038/s41534-020-0272-6}
C.~Blank, D.~K. Park, J.-K.~K. Rhee, and F.~Petruccione, ``{Quantum classifier
  with tailored quantum kernel},'' \emph{npj Quantum Information}, vol.~6,
  no.~1, p.~41, 2020.

\bibitem{Aaronson2015}
S.~Aaronson, ``Read the fine print,'' \emph{Nature Physics}, vol.~11, pp.
  291--293, 2015.

\bibitem{Leymann}
F.~Leymann and J.~Barzen, ``The bitter truth about gate-based quantum
  algorithms in the {NISQ} era,'' \emph{Quantum Science and Technology},
  vol.~5, no.~4, p. 044007, 2020.

\bibitem{bergholm2005quantum}
V.~Bergholm, J.~J. Vartiainen, M.~M{\"o}tt{\"o}nen, and M.~M. Salomaa,
  ``Quantum circuits with uniformly controlled one-qubit gates,''
  \emph{Physical Review A}, vol.~71, no.~5, p. 052330, 2005.

\bibitem{Plesch2011}
M.~Plesch and {\v{C}}.~Brukner, ``Quantum-state preparation with universal gate
  decompositions,'' \emph{Physical Review A}, vol.~83, no.~3, p. 032302, 2011.

\bibitem{malvetti2021quantum}
E.~Malvetti, R.~Iten, and R.~Colbeck, ``Quantum circuits for sparse
  isometries,'' \emph{Quantum}, vol.~5, p. 412, 2021.

\bibitem{Araujo0407}
I.~F. Araujo, D.~K. Park, F.~Petruccione, and A.~J. da~Silva, ``A
  divide-and-conquer algorithm for quantum state preparation,''
  \emph{Scientific Reports}, vol.~11, p. 6329, 2021.

\bibitem{araujo2021configurable}
I.~F. Araujo, D.~K. Park, T.~B. Ludermir, W.~R. Oliveira, F.~Petruccione, and
  A.~J. Da~Silva, ``\BIBforeignlanguage{en}{Configurable sublinear circuits for
  quantum state preparation},'' \emph{\BIBforeignlanguage{en}{Quantum
  Information Processing}}, vol.~22, no.~2, p. 123, 2023.

\bibitem{zhang2021low}
X.-M. Zhang, M.-H. Yung, and X.~Yuan, ``Low-depth quantum state preparation,''
  \emph{Phys. Rev. Research}, vol.~3, p. 043200, 2021.

\bibitem{Park2019}
D.~K. Park, F.~Petruccione, and J.-K.~K. Rhee, ``Circuit-based quantum random
  access memory for classical data,'' \emph{Scientific Reports}, vol.~9, p.
  3949, 2019.

\bibitem{zoufal2019quantum}
C.~Zoufal, A.~Lucchi, and S.~Woerner, ``Quantum generative adversarial networks
  for learning and loading random distributions,'' \emph{npj Quantum
  Information}, vol.~5, p. 103, 2019.

\bibitem{nakaji2021}
K.~Nakaji, S.~Uno, Y.~Suzuki, R.~Raymond, T.~Onodera, T.~Tanaka, H.~Tezuka,
  N.~Mitsuda, and N.~Yamamoto, ``Approximate amplitude encoding in shallow
  parameterized quantum circuits and its application to financial market
  indicator,'' 2021.

\bibitem{javier2021}
G.~Marin-Sanchez, J.~Gonzalez-Conde, and M.~Sanz, ``Quantum algorithms for
  approximate function loading,'' 2021.

\bibitem{mozafari2021efficient}
F.~Mozafari, H.~Riener, M.~Soeken, and G.~De~Micheli, ``Efficient boolean
  methods for preparing uniform quantum states,'' \emph{IEEE Transactions on
  Quantum Engineering}, vol.~2, pp. 1--12, 2021.

\bibitem{mozafari_2022}
F.~Mozafari, G.~De~Micheli, and Y.~Yang, ``Efficient deterministic preparation
  of quantum states using decision diagrams,'' \emph{Phys. Rev. A}, vol. 106,
  p. 022617, 2022.

\bibitem{gleinig_2021}
N.~Gleinig and T.~Hoefler, ``An efficient algorithm for sparse quantum state
  preparation,'' in \emph{2021 58th ACM/IEEE Design Automation Conference
  (DAC)}, 2021, pp. 433--438.

\bibitem{veras2021}
T.~M.~L. de~Veras, L.~D. da~Silva, and A.~J. da~Silva, ``Double sparse quantum
  state preparation,'' 2021.

\bibitem{MORA2006}
C.~E. Mora and H.~J. Briegel, ``Algorithmic complexity of quantum states,''
  \emph{International Journal of Quantum Information}, vol.~04, pp. 715--737,
  2006.

\bibitem{PhysRevLett.95.200503}
------, ``Algorithmic complexity and entanglement of quantum states,''
  \emph{Phys. Rev. Lett.}, vol.~95, p. 200503, 2005.

\bibitem{PhysRevLett.86.5188}
R.~Raussendorf and H.~J. Briegel, ``A one-way quantum computer,'' \emph{Phys.
  Rev. Lett.}, vol.~86, pp. 5188--5191, 2001.

\bibitem{Barreiro2010}
J.~T. Barreiro, P.~Schindler, O.~Gühne, T.~Monz, M.~Chwalla, C.~F. Roos,
  M.~Hennrich, and R.~Blatt, ``Experimental multiparticle entanglement dynamics
  induced by decoherence,'' \emph{Nature Physics}, vol.~6, pp. 943--946, 2010.

\bibitem{PhysRevA.54.1098}
A.~R. Calderbank and P.~W. Shor, ``Good quantum error-correcting codes exist,''
  \emph{Phys. Rev. A}, vol.~54, pp. 1098--1105, 1996.

\bibitem{Steane1996}
A.~Steane, ``Multiple-particle interference and quantum error correction,''
  \emph{Proceedings of the Royal Society of London. Series A: Mathematical,
  Physical and Engineering Sciences}, vol. 452, pp. 2551--2577, 1996.

\bibitem{PhysRevLett.83.648}
R.~Cleve, D.~Gottesman, and H.-K. Lo, ``How to share a quantum secret,''
  \emph{Phys. Rev. Lett.}, vol.~83, pp. 648--651, 1999.

\bibitem{Cirac2000}
W.~Dür and J.~I. Cirac, ``Multiparty teleportation,'' \emph{Journal of Modern
  Optics}, vol.~47, no. 2-3, pp. 247--255, 2000.

\bibitem{shende2006synthesis}
V.~V. Shende, S.~S. Bullock, and I.~L. Markov, ``Synthesis of quantum-logic
  circuits,'' \emph{IEEE Transactions on Computer-Aided Design of Integrated
  Circuits and Systems}, vol.~25, no.~6, pp. 1000--1010, 2006.

\bibitem{Iten2016}
R.~Iten, R.~Colbeck, I.~Kukuljan, J.~Home, and M.~Christandl, ``Quantum
  circuits for isometries,'' \emph{Physical Review A}, vol.~93, no.~3, p.
  032318, mar 2016.

\bibitem{qclib}
I.~F. Araujo, C.~Blank, A.~da~Silva, I.~Cesar, and L.~Silva, ``Quantum
  computing library (qclib),'' \url{https://github.com/qclib/qclib}, 2022.

\bibitem{Qiskit}
{Qiskit contributors}, ``Qiskit: An open-source framework for quantum
  computing,'' 2023.

\bibitem{blank2021quantum}
C.~Blank, D.~K. Park, and F.~Petruccione, ``Quantum-enhanced analysis of
  discrete stochastic processes,'' \emph{npj Quantum Information}, vol.~7,
  no.~1, pp. 1--9, 2021.

\bibitem{10.1007/s11128-017-1633-8}
P.~Teng, ``{Accurate calculation of the geometric measure of entanglement for
  multipartite quantum states},'' \emph{Quantum Information Processing},
  vol.~16, no.~7, p. 181, 2017.

\bibitem{mehlhorn_algorithms_2008}
K.~Mehlhorn and P.~Sanders, \emph{Algorithms and data structures: the basic
  toolbox}.\hskip 1em plus 0.5em minus 0.4em\relax Berlin: Springer, 2008.

\bibitem{skiena_algorithm_2008}
S.~S. Skiena, \emph{The algorithm design manual}, 2nd~ed.\hskip 1em plus 0.5em
  minus 0.4em\relax London: Springer, 2008, oCLC: ocn228582051.

\bibitem{Nemhauser1978}
G.~L. Nemhauser, L.~A. Wolsey, and M.~L. Fisher, ``An analysis of
  approximations for maximizing submodular set functions—i,''
  \emph{Mathematical Programming}, vol.~14, pp. 265--294, 1978.

\bibitem{cormen_introduction_2009}
T.~H. Cormen, Ed., \emph{Introduction to algorithms}, 3rd~ed.\hskip 1em plus
  0.5em minus 0.4em\relax Cambridge, Mass: MIT Press, 2009, oCLC: ocn311310321.

\bibitem{10.1111/j.1749-6632.1995.tb39008.x}
A.~Shimony, ``{Degree of Entanglement},'' \emph{Annals of the New York Academy
  of Sciences}, vol. 755, no.~1, pp. 675--679, 1995.

\bibitem{10.1088/0305-4470/34/35/305}
H.~Barnum and N.~Linden, ``{Monotones and invariants for multi-particle quantum
  states},'' \emph{Journal of Physics A: Mathematical and General}, vol.~34,
  no.~35, p. 6787, 2001.

\bibitem{10.1103/physreva.68.042307}
T.-C. Wei and P.~M. Goldbart, ``{Geometric measure of entanglement and
  applications to bipartite and multipartite quantum states},'' \emph{Physical
  Review A}, vol.~68, no.~4, p. 042307, 2003.

\bibitem{Chakraborty2015}
S.~Chakraborty, ``Generating discrete analogues of continuous probability
  distributions-a survey of methods and constructions,'' \emph{Journal of
  Statistical Distributions and Applications}, vol.~2, p.~6, 2015.

\bibitem{KOGBETLIANTZ_1955}
E.~G. KOGBETLIANTZ, ``Solution of linear equations by diagonalization of
  coefficients matrix,'' \emph{Quarterly of Applied Mathematics}, vol.~13,
  no.~2, pp. 123--132, 1955.

\bibitem{forsythe_1960}
G.~E. Forsythe and P.~Henrici, ``The cyclic jacobi method for computing the
  principal values of a complex matrix,'' \emph{Transactions of the American
  Mathematical Society}, vol.~94, no.~1, pp. 1--23, 1960.

\bibitem{10.1103/revmodphys.90.015002}
T.~Albash and D.~A. Lidar, ``{Adiabatic quantum computation},'' \emph{Reviews
  of Modern Physics}, vol.~90, no.~1, p. 015002, 2018.

\bibitem{10.1063/1.4880755}
L.~Veis and J.~Pittner, ``{Adiabatic state preparation study of methylene},''
  \emph{The Journal of Chemical Physics}, vol. 140, no.~21, p. 214111, 2014.

\end{thebibliography}

\end{document}